\documentclass[iop, apj]{emulateapj}

\usepackage{epsf}
\usepackage{color}
\usepackage{hyperref}
\hypersetup{
pdftitle={},
pdfauthor={Vasyl Yurchyshyn},     
pdfsubject={Scientific Paper},   
pdfkeywords={sunspot umbra} {oscillations} {chromosphere} {jets}, 
colorlinks=true,       
linkcolor=black,       
citecolor=black,       
filecolor=black,
urlcolor=blue
}

\newcommand{\comments}[1]{}

\begin{document}


\title{High Resolution Observations of Chromospheric Jets in Sunspot Umbra}

\author{Yurchyshyn, V., Abramenko, V., Kosovichev, A., and Goode, P.}

\affil{\it Big Bear Solar Observatory, New Jersey Institute of Technology, Big Bear City, CA 92314, USA}

\begin{abstract}

Recent observations of sunspot's umbra suggested that it may be finely structured at a sub-arcsecond scale representing a mix of hot and cool plasma elements. In this study we report the first detailed observations of the umbral spikes, which are cool jet-like structures seen in the chromosphere of an umbra. The spikes are cone-shaped features with a typical height of 0.5-1.0~Mm and a width of about 0.1~Mm. Their life time ranges from 2 to 3 ~min and they tend to re-appear at the same location. The spikes are not associated with photospheric umbral dots and they rather tend to occur above darkest parts of the umbra, where magnetic fields are strongest. The spikes exhibit up and down oscillatory motions and their spectral evolution suggests that they might be driven by upward propagating shocks generated by photospheric oscillations. It is worth noting that triggering of the running penumbral waves seems to occur during the interval when the spikes reach their maximum height.

\end{abstract}

\comments{Key words: Sun: atmosphere -- Sun: chromosphere -- Sun: oscillations}

\section{Introduction}

\noindent Recent advancements in observations and modeling created a picture of a sunspot as being very dynamic and complex magnetic structure \citep[e.g.,][]{2012ApJ...747L..18S, 2011Sci...333..316S, 2011LRSP....8....3R,2011ApJ...740...15R}. A closer look at the dark sunspot umbra using photospheric spectral lines revealed detailed structure of bright and nearly circular small intensity patches called umbral dots \citep[UDs,][and references therein]{2012ApJ...745..163K, 0004-637X-752-2-109}. Magnetic fields in UDs are weaker than in their darker surroundings and UDs show plasma up-flows of a few hundred m s$^{-1}$ \citep{1993A&A...278..584W, 2004ApJ...614..448S, 2004ApJ...604..906R, 2008ApJ...672..684R,2009ApJ...702.1048W, 2012ApJ...757...49W}. According to realistic 3D simulations \citep[e.g.][]{2006ApJ...641L..73S, Bharti_2011} UDs result from magneto-convection in sunspot umbra, and represent narrow convective up-flow plumes with adjacent down-flows, which become almost field-free near the surface layer. 

The umbra appears much more dynamic when observed using chromospheric spectral lines.  Based on recent high resolution observations it has been suggested that the sunspot's umbra may be finely structured, and consists of hot and cool plasma elements intermixed at sub-arcsecond scales \citep{2000Sci...288.1396S,2009ApJ...696.1683S, 2013ApJ...776...56R}. The well-known 3-min sunspot oscillations \citep[e.g.,][and references therein]{2000SoPh..192..373B,2013SoPh..288...73M} often lead to the appearance of bright umbral flashes (UFs), which are emissions in the core of chromospheric lines caused by hot shocked plasma \citep{2010ApJ...722..888B} and were reported to display a filamentary structure \citep{2009ApJ...696.1683S}. Later studies also found evidence for a two-component structure of the umbra \citep{2005ApJ...635..670C,2013A&A...556A.115D,2013ApJ...776...56R}. Transient jet-like structures were recently reported by \cite{2013A&A...552L...1B} seen in the Ca II H images of the sunspot umbrae, which they called umbral microjets. The microjets appear to be aligned with the umbral field, and no one-to-one correspondence between the microjets and the umbral dots was found. \cite{2013ApJ...776...56R} described small-scale, periodic, jet-like features in the chromosphere above sunspots, which result from long-period waves leaking into the chromosphere along inclined sunspot fields.

In this paper we present first detailed observations of fine-scale chromospheric phenomena in sunspot's umbra using H$\alpha$ imaging spectroscopy data obtained with the New Solar Telescope \cite[NST,][]{goode_apjl_2010} that operates in Big Bear Solar Observatory. The data allowed us to fully resolve ubiquitous dynamic umbral H$\alpha$ jet-like features and measure their general properties.

\begin{figure*}[!th]
\centering
\begin{tabular}{c}
\epsfxsize=5.0truein  \epsffile{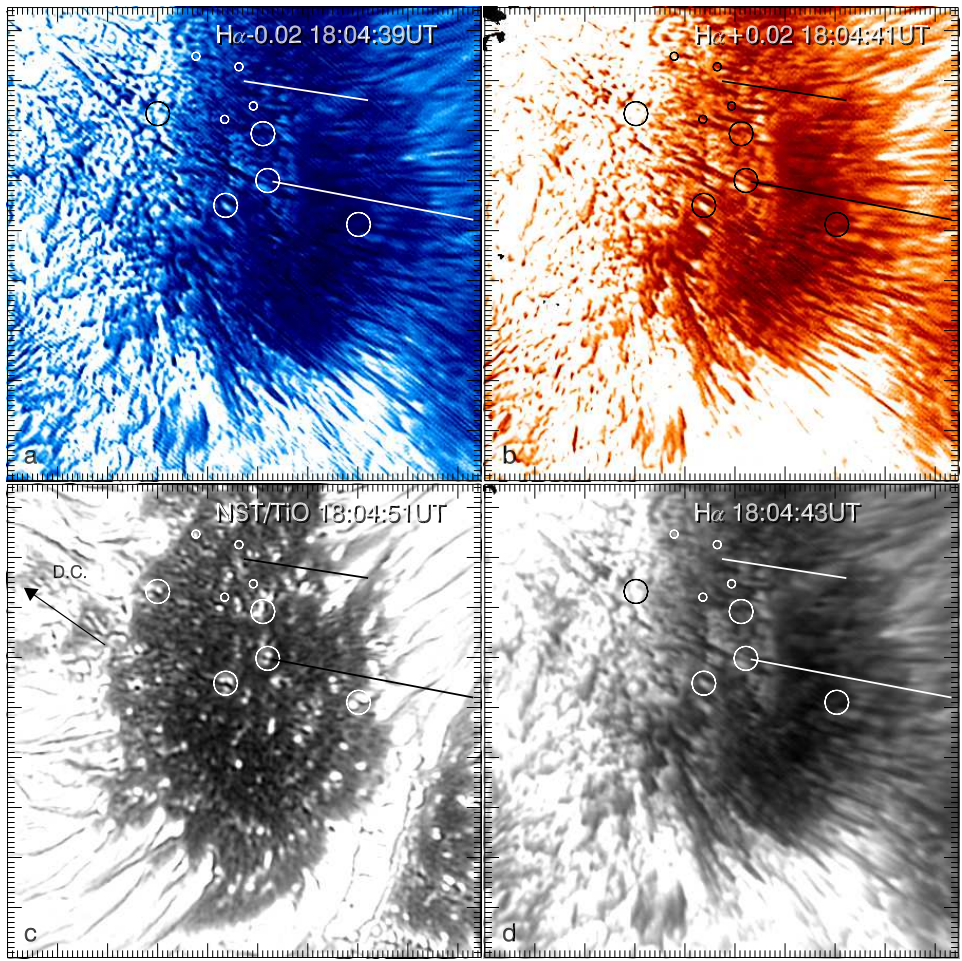} \
\end{tabular}
\caption{The main spot of NOAA AR 11768 as seen in the chromospheric H$\alpha$ line (top and lower right panels) as well as the photospheric TiO 705~nm line (lower left). The top row shows off-band images of the sunspot taken at -0.02~nm (left, blue) and +0.02~nm (right, red) off the line center. All images were unsharp masked. The D.C. arrow in the lower left image points toward the disk center. The two line segments indicate the location of \textit{xt} cuts. The large circles indicate several bright photospheric umbral dots with fine structures inside them. The small circles mark the base of several spikes (see text for more details). The long tick marks separate 1~Mm intervals. The images were not corrected for the projection effect.}
\label{sunspot}
\end{figure*}

\section{Data}

\noindent On 14 June 2013 the NST was pointed at the leading spot of active region NOAA 11768 located at S11W46 (Figure \ref{sunspot}). Simultaneous observations were performed for the titanium oxide (TiO) band using the Broadband Filter Imager (BFI) and at the H$\alpha$ spectral line using the Visible Imaging Spectrometer (VIS). The photospheric images of the sunspot were acquired every 25~s using a 1~nm passband TiO filter centered at 705.7~nm with the pixel scale of 0''.0375. This absorption line (the head of the TiO $\gamma$-system) is only formed at low temperatures below 4000~K, i.e., inside sunspots \citep[see Fig. 10][]{2003A&A...412..513B}.

The VIS combines a 0.5~nm interference filter with a Fabry-P\'{e}rot etalon to produce a bandpass of 0.007~nm over a 70'' field of view centered at the H$\alpha$ spectral line. Imaging of the chromosphere were performed at 11 position along the spectral line with a 0.02~nm step along the spectrum and the pixel size of 0''.029. The difference in the acquisition time at two sequential line positions (e.g, +0.02~nm and -0.02~nm) was about 2~s. The 11 point line scan was recorded every 25~s along the following sequence: -0.1, +0.1, -0.08, +0.08, -0.06, +0.06, -0.04, +0.04, -0.02, +0.02, 0.0~nm. At each line position we acquired a burst of 25 images with the exposure times ranging from 7~ms (at -0.1~nm) to 25~ms (at the line center). These bursts were used for speckle reconstruction, so that each line position was speckled separately. To estimate an error in the line profile we used the 50 profiles shown in the last panel of Figure \ref{lambdat}. Each profile was constructed from a series of 11 speckle reconstructed off-band H$\alpha$ images (see above), so that the errors introduced by local misalignment and residual seeing are taken into account. These profiles were measured inside of a stable, slowly evolving super-penumbra canopy, so that we accept that all variations in the line profiles are due to varying seeing as well as errors introduced by the speckle reconstruction code. We calculated relative standard deviation using 50 data points at each line position and found that it ranges from 2\% to 9\% with the average of 5\%.

All images were acquired with the aid of an adaptive optics (AO) system, which incorporates a 357 actuator deformable mirror, a Shack-Hartmann wavefront sensor with 308 sub-apertures, and an digital signal processor system. The Kiepenheuer-Institut f{\"u}r Sonnenphysik's software package for speckle interferometry of adaptive optics corrected solar data \citep[KISIP,][]{kisip_code} was applied to all acquired images as a post AO reconstruction algorithm allowing us to achieve the diffraction limit of the telescope over a large FOV.

\section{Results}

\noindent In Figure \ref{sunspot} we show constrast-enhanced (unsharp masked) H$\alpha$-0.02~nm (top left),  H$\alpha$+0.02~nm (top right), H$\alpha$ (lower right) and TiO (lower left) images of the sunspot. The dynamics of the umbral area is best seen in the  H$\alpha$-0.02~nm movie
\href{http://www.bbso.njit.edu/~vayur/spikes/S1.mp4}{(S1.mp4)}\color{black}, where the diffuse, dark and dynamic intensity fronts are seen to rapidly travel across the umbra, seemingly reflecting back and forth from the umbra-penumbra boundary triggering penumbral waves. The movie images were not processed with the unsharp masking method. The chromospheric ‘’surface’’ of the umbra seems to be vertically oscillating giving an impression of standing surface waves in an enclosed body of fluid. The bright diffuse patches seen within the umbra perimeter around the three central mid-size circles are UFs. At the spatial resolution of $\sim$ 0''.1 the UFs do not show any intrinsic fine structure. Instead, the bright patches appear to be interrupted by vertical dark spikes projected on the surrounding UFs background.

\begin{figure}[!th]
\centering
\begin{tabular}{c}
\epsfxsize=3.0truein  \epsffile{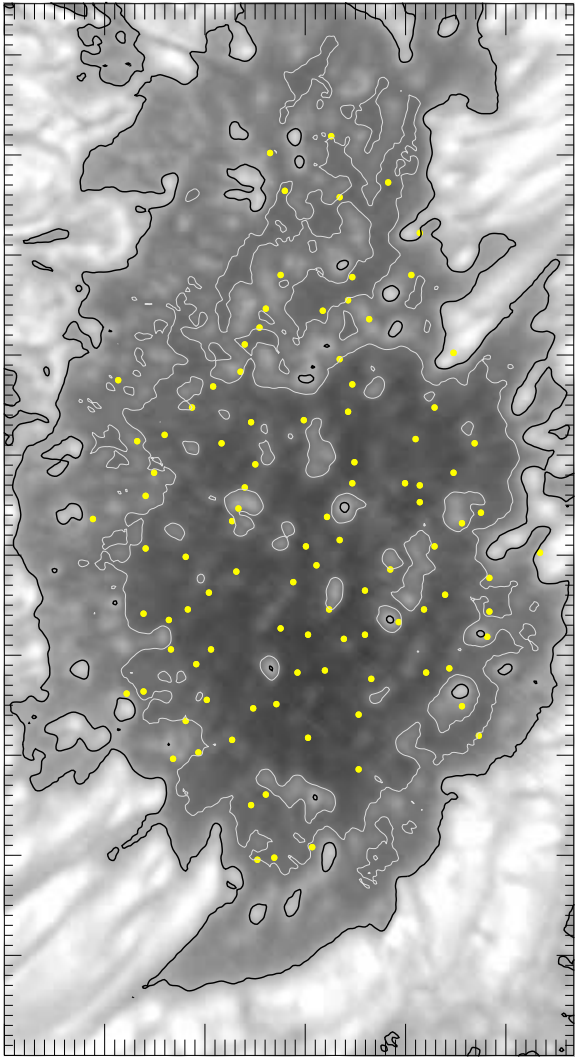} \
\end{tabular}
\caption{ (left) Positions of the base of umbral spikes (yellow dots) plotted over an H$\alpha$-0.1~nm image. The black contour is at the intensity level of 1.5$\times$10$^{3}$DN and outlines the boundary of the umbra. The white contour is at 1.0$\times$10$^{3}$DN intensity level and it corresponds to the peak in the umbra intensity distribution (see Fig. \ref{hist}).}\label{base}
\end{figure}

According to Figure \ref{sunspot} these umbral spikes can be detected everywhere within the umbra with nearly the same number density, although it seems that they preferably occur in darker parts of the umbra. Using the large circles we indicate in the TiO image several bright photospheric UDs that have dark lanes and/or dots inside them \citep[see][]{2008ApJ...672..684R}. Comparison with the chromospheric images shows that in 4 cases out of 5 an umbral spike can be found in the vicinity of an UD. It is possible that the surrounding features simply overlap with the UDs, since we did not find strong evidence in those cases for the spikes to originate above UDs.  We also plotted small circles at the base of several spikes in the off-band images and it appears that all of them map back to dark areas in the photospheric umbra.

\begin{figure}[!th]
\centering
\begin{tabular}{c}
\epsfxsize=3.0truein  \epsffile{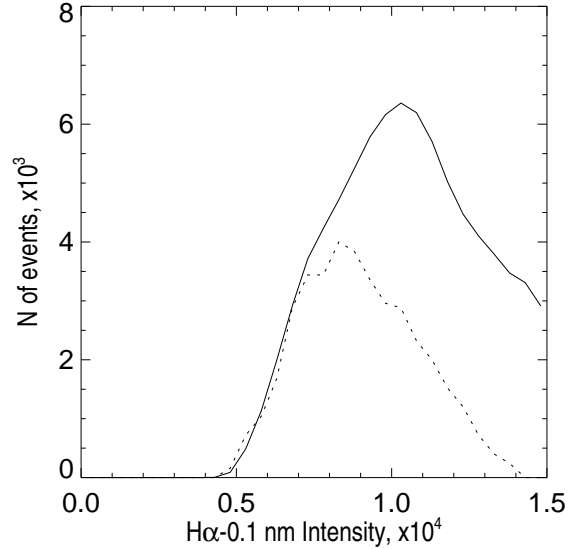} \
\end{tabular}
\caption{(right) Distribution of umbra brightness (solid line) determined for the areas enclosed by the black contour in Fig. \ref{base}. The distribution peaks at the 1.0$\times$10$^{3}$DN intensity level marked by the white contour in Fig. \ref{base}. The dotted line shows the base intensity distribution, i.e., umbral intensity measured at the base of umbral spikes. The base intensity distribution was multiplied by factor of 400 to make these two curves comparable.}\label{hist}
\end{figure}

In Figure \ref{base} the position of the base of umbral spikes (yellow dots) is plotted over an image of the umbra as observed at H$\alpha$-0.1~nm. There were total 101 spikes identified in this image and the position of their base was determined manually using the H$\alpha$-0.02~nm image shown in Figure \ref{sunspot}. This plot and Figure \ref{base} further show that the spikes tend avoid bright UDs and are either co-spatial with dark umbral patches or are found in the vicinity of an UD. That probably explains presence of spikes near large UDs. In Figure \ref{hist} we plot the brightness distribution (solid line) determined for the areas enclosed by the black contour plotted at the intensity level of 1.5$\times$10$^{3}$DN. The distribution peaks at the 1.0$\times$10$^{3}$DN intensity level marked by the white contour line. The dotted line shows the base intensity distribution measured under the base of umbral spikes. In this plot, the base intensity distribution was multiplied by factor of 400 to make these two curves comparable. The distributions clearly show the preference for spikes to originate in the dark umbral patches.

\begin{figure*}[!t]
\centering
\begin{tabular}{c}
\epsfxsize=5.0truein  \epsffile{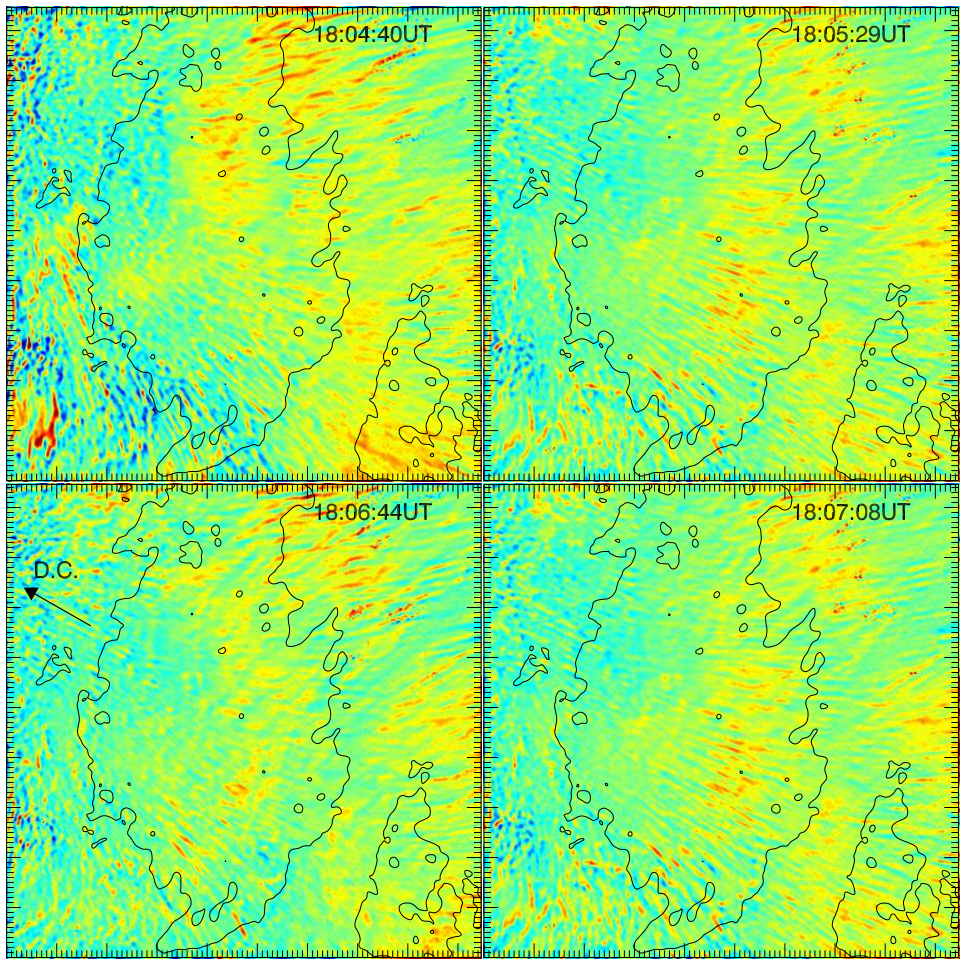} \
\end{tabular}
\caption{H$\alpha$ 0.02~nm Doppler maps. The blue/red colors indicate up and down flows. The size of field of view is the same as in Figure \ref{sunspot} and was shifted to the right by about 1.6~Mm. The thin solid contour outlines the boundary of the umbra. The long tick marks separate 1~Mm intervals.}
\label{dopp}
\end{figure*}

The spikes have a cone-like appearance with a wide base (typically $\sim$0.1~Mm) and a pointed tip. Their typical length is between 0.5 and 1.0~Mm. Using the off-band images in Figure \ref{sunspot} and the line scan data (see \href{http://www.bbso.njit.edu/~vayur/spikes/S2.mp4}{S2.mp4} \color{black} movie) we conclude that i) umbral spikes simultaneously show up in the blue and red wings of the H$\alpha$ spectral line, ii) their number density rapidly decreases when observed further away from the line center and iii) they show oscillatory up and down motions. When observing near the H$\alpha$ line center the umbra is seen filled with chromospheric plasma forming various extended structures, so that the spikes are less prominent and only their dark tips can be identified when carefully comparing off-band and line center images. 

\cite{2012ApJ...749..136L} performed MHD simulations and three-dimensional (3D) non-LTE radiative transfer computations to understand details of the H$\alpha$ line formation. They reported that H$\alpha$ opacity in the upper chromosphere is mainly sensitive to the mass density and only weakly sensitive to the temperature. Also, the intensity of the H$\alpha$ line-core is related to the formation height and the intensity decreases as the formation height is shifted toward the upper chromosphere. In their simulations the fibril like H$\alpha$ structures represent ridges of enhanced mass density, which displace the formation level of the H$\alpha$ line-core to the higher levels, so these structures appear darker. Based on these simulation results, we interpret umbral spikes as plasma structures with enhanced mass density.

In Figure \ref{dopp} we plot Doppler-shift maps obtained by subtracting red wing H$\alpha$+0.02~nm images from the corresponding blue-shifted images taken 2~s prior. In general, the Doppler-shift maps are finely structured but do not show presence of any strong directed out- or down-flows associated with the spikes. Instead, they  are seen as very low contrast Doppler-shift features. The red-shifted Doppler features seen in these maps at the limb side of the umbra and penumbra, appear to be longer (1-2~Mm) and they are mostly seen in the projection on the sunspot's penumbra. The sunspot was located at a longitude of 45$\deg$ west, therefore, near the disk side umbra-penumbra boundary, where the field inclination is nearly 45$\deg$ the line-of-sight is approximately parallel to the field lines, while at the limb side of the boundary, the line of sight is nearly perpendicular to the magnetic field lines. Thus, because of the projection effect, the spikes at the limb side are more easily observed as compared to their point like projection at the disk side of the umbra.  

\begin{figure}[!t]
\centering
\begin{tabular}{c}
\epsfxsize=3.3truein  \epsffile{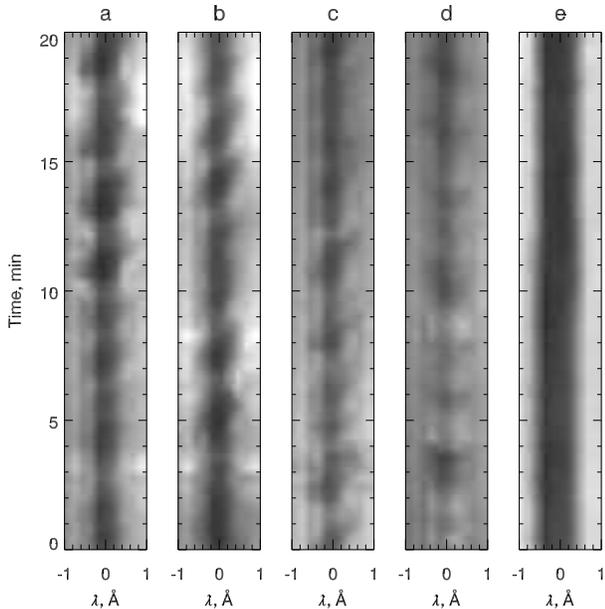} \
\end{tabular}
\caption{Examples of spectral evolution as measured for 2 different umbral spikes (panels \textit{a-b}). Panel \textit{c} and \textit{d} show spectral evolution immediately next to a spike and in UFs, correspondingly. Panel \textit{e} represents spectral evolution in a super-penumbral filament outside the sunspot. 50 line profiles acquired within a 20~min time span were used to produce these plots.}
\label{lambdat}
\end{figure}

The four Doppler maps in Figure \ref{dopp} illustrate that the location of the red-shifted thorns within the umbra varies rapidly in time, which is consistent with the idea of their wave origin. The observed behavior suggests that the spikes may have a wave origin rather than being linear jet-like outflows from the sunspot’s umbra. As it follows from the maps, the intensity of the red-shifted fine structures evolve in time, however, no equally strong blue shifted signal appears instead as would be expected from simple oscillatory motions. The distribution patterns of these red-shifted features varies rapidly in time, which is consistent with the idea of their wave origin. We therefore interpret these jets as the outflows in the sunspot dynamic fibrils described in \cite{2013ApJ...776...56R}.

In Figure \ref{lambdat} we show an $\lambda\,t$ plot representing the time evolution of the H$\alpha$ line profiles measured at various locations inside the umbra. Panel \textit{e} is shown for a comparison and it represents a co-temporal spectral evolution measured in a super-penumbral filament outside the sunspot. The \textit{a-b} profile stacks were measured inside the spikes and they show typical signatures of а shock driven flows \citep{2006ApJ...647L..73H,2008ApJ...673.1194L} , i.e., the initially blue-shifted spectral line gradually becomes red-shifted due to deceleration and the subsequent fall of previously upwardly moving chromospheric plasma (see, e.g., panel \textit{b} between $ t=13 $ and $ t=17 $~min). The diagonal spectral features seem to repeat approximately every 2-3~min, although the line intensity and the spectral shift varies over time and from one location to another. Panel \textit{c} shows the spectral line evolution measured immediately next to the spikes. The shocks are present there too, although the plasma seems to be compressed to a lesser degree as evidenced by the much shallower line profiles. Panel \textit{d} shows similar data but for pixels belonging to UFs. The diagonal shock features are largely absent in this case. Instead, periods of emission (e.g., at t=8-9~min) and absorption (e.g. at t=6-5~min) follow each other with periodicity of about 2-3~min.

\begin{figure*}[th!]
\centering
\begin{tabular}{c}
\epsfxsize=7.0truein  \epsffile{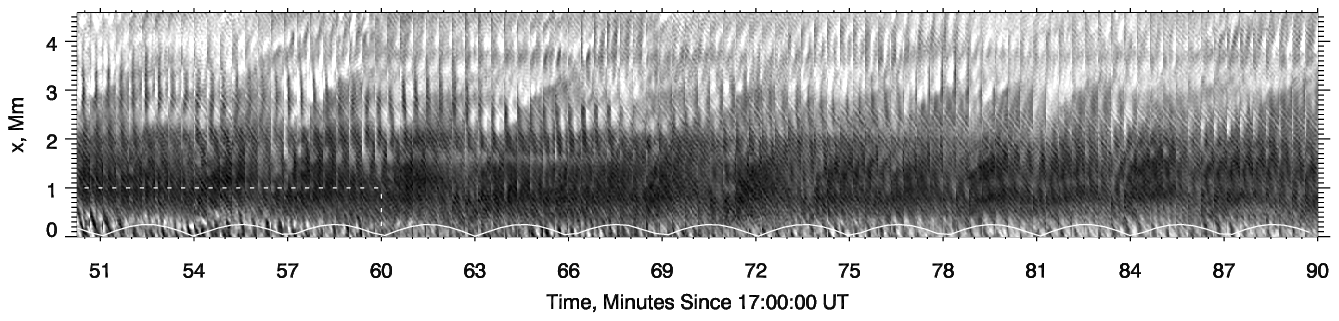} \\
\epsfxsize=7.0truein  \epsffile{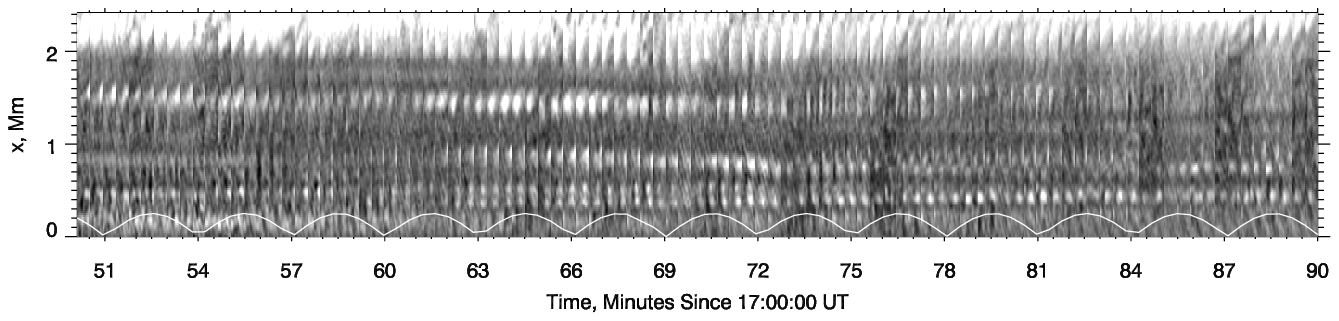} \
\end{tabular}
\caption{Top panel: Time-distance plots showing chromospheric dynamics as it is seen at the H$\alpha$-0.02~nm along the longer slit in Figure \ref{sunspot}. The width of the slit was chosen to be 0.32~Mm. The white curve in both panels has a 3~min period and is plotted for illustration only. The white box outlines an area enlarged in Fig. \ref{xtplot_inset}. The bottom panel shows similar dynamics, however, observed at the H$\alpha$-0.06~nm along the shorter slit.} \label{xtplot}
\end{figure*}

\begin{figure}[!th]
\centering
\epsfxsize=3truein  \epsffile{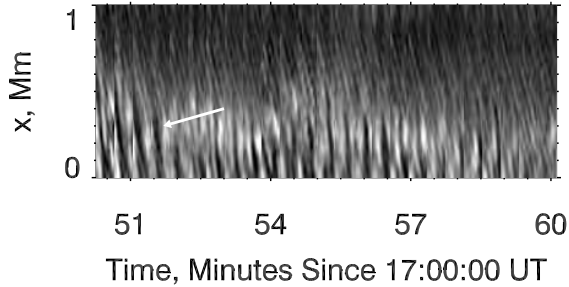}
\caption{Enlarged fragment of the \textit{xt} plot shown in the top panel of Fig. \ref{xtplot}. Arrows indicate umbral spikes.}
\label{xtplot_inset}
\end{figure}

In Figures \ref{xtplot} and \ref{xtplot_inset} we show time variations of chromospheric intensity made along the cuts indicated in Figure \ref{sunspot} by two line segments. The cuts were made along the axis of spikes and the width of the slit was 0.32~Mm (15~pixels=0''.43). The repeating dark half-parabola feature seen above 1~Mm is a propagating front of penumbral waves that are excited at the umbra-penumbra boundary (x=1~Mm) and travel toward the outer bounds of the penumbra. The umbral spikes seen here as the dense and short features visible below $x<0.5$~Mm and between $ t=50 $ and $ t=58 $~min (Figure \ref{xtplot_inset}) with the lifetime of about 3~min. The tip of the spikes appears to be rising up with velocities of about 4-5~km s$^{-1}$, which is expected to reflect local sound speed and is in the range of the reported velocities associated with oscilaltions and shock formation \cite[e.g.,][]{1998ApJ...495..468S, 2013SoPh..288...73M}, the temporal resolution of the data did not allow us to make a more detailed conclusions about their dynamics. After $ t=60 $~min they briefly appear nearly every 3~min approximately at the moments of the minimum in the white curve (i.e, at $ t=66, 69, 72, ... 84, $ and $ 87 $~min). In general, the data set gives an impression that the spikes tend to re-occur at the same locations at fairly regular time intervals; however, their parameters vary, possibly reflecting the strength of local oscillations. The lower plot in Figure \ref{xtplot} shows similar features and behavior except that it was made using H$\alpha$-0.06~nm data. The penumbral waves are still present, although their front is less dense and narrower. The same is true about the umbral spikes, which are only weakly seen in the plot (i.e, at $ t= 65, 67, 73, 76 $, and $ 79 $~min). 

It is worth noting that each occurrence of the spikes in the center of the penumbra seems to be correlated in time with the onset of a new penumbral wave: the wave begins when spikes are at the peak of their development and is best evidenced in the lower panel of Figure \ref{xtplot} between t=63~min and t=81~min. The spikes are visible under the white curve and the middle of their life time occurs when the white curve is in the maximum. At the same moment of time a dark front, representing penumbral waves, begins to propagate above the white curve. This correlation may indicate that the spikes and the running penumbral waves are triggered and powered by the same process that appear to be coherent on scales comparable to the size of a sunspot, such as sunspot oscillations.

\section{Conclusions}

\noindent 

We summarize our findings as follows. High resolution NST observations of solar chromosphere using the H$\alpha$ spectral line revealed the existence of dynamic spike-like chromospheric structures inside sunspot's umbra, which we called umbral spikes. The spikes are on average about 0.1~Mm wide, their height does not exceed 1~Mm. They are mainly vertical with a slight tendency to fan out closer to the periphery of the umbra. Since umbral fields show similar inclination distribution we suggest that the umbral spikes are aligned with the umbral magnetic fields. The spikes show a nearly uniform distribution over the umbra with a tendency to be more concentrated in its darkest parts occupied by strongest fields in the umbra, while UDs are considered to be field-free magneto-convection features. Thus, our finding seem to indicate that the umbral spikes may be co-spatial with strong magnetic flux concentrations rather than with the weaker magnetic fields above UDs. 

These first detailed observations of umbral spikes presented here suggest that they are a wave phenomena and result from sunspot oscillations. \cite{2013A&A...552L...1B} observed transient jet-like structures in Ca II H images of sunspot umbrae, which they called umbral microjets. The authors speculated that the microjets, which are shorter than 1~Mm and not wider than 0''.3, may be either upflow jets driven by the pressure gradient above the photospheric UDs or they may be caused by reconnection of hypothetical opposite polarity fields that might exist around large UDs. Although the H$\alpha$ spikes and the microjets are of a comparable size, the lifetime of spikes appears to be longer (2-3~min) than that of the microjets (50~s). While a possible relationship between them has yet to be established, it seems that the two phenomena are different and the scenarios presented in \cite{2013A&A...552L...1B} may not be able to explain the spikes, since the latter show a preference to be more frequent in the darkest cores of the umbra, dominated by strong fields and void of large and bright UDs.

\cite{2013ApJ...776...56R} described dynamic fibrils (DF) observed in the chromosphere above a sunspot. Their \textit{xt} plot generated from a series of the H$\alpha$ line center images \citep[see Fig. 3 in][]{2013ApJ...776...56R} shows the presence of parabolic intensity features, which were interpreted by these authors as being very short jet-like features precisely above the umbra. At the same time individual umbral spikes where not resolved in their data and only diffuse, low contrast dark specks can be distinguished inside the umbra in the corresponding H$\alpha$ images. The umbral spikes and the DFs may be the same wave phenomenon while their differences in appearance are caused by the fine scaled structures in magnetic fields, viewing angle as well as the resolution of the data.

\cite{2011ApJ...743..142H} studied wave propagation in different magnetic configurations. They found that peaks of power of 3~min oscillations and high amplitudes of vertical velocities (5-8~km/s) are located above strong photospheric flux concentrations or, in other words, inside vertical flux tubes. They also found that rising and falling jets, which form as a result of the oscillations, have their axis aligned with the magnetic axis of field concentrations. General appearance of the simulated jets and their oscillatory motions are strikingly similar to those of the spikes (see Fig. 25 in \cite{2011ApJ...743..142H} and Fig. 1 in this paper), although the simulated jets appear to be, on average, more extended (0.5-6.0~Mm), longer living (2-5~min) and show higher velocities (10-40~km/s). Thus one possible interpretation of the observed phenomena is the penetration of photospheric oscillations into the chromosphere along thin and vertical magnetic flux tubes.

Using spectropolarimetric observations \cite{2000Sci...288.1396S} suggested the existence of an unresolved active component with upward directed velocities. More precisely, the anomalous polarization profiles could only be explained by emission from an unresolved mixture of upward propagating shock and a cool slowly downflowing surroundings. Later \cite{2005ApJ...635..670C} specified that the active component is present through out the entire oscillation cycle. They also inferred that the shock waves propagate into the umbra inside channels of subarcsecond width, which could be the flux tubes discussed above. Finally, \cite{2009ApJ...696.1683S} presented high resolution \textit{Hinode} data on UFs concluding that UFs show fine filamented structure. 

We argue that the 0''.1 wide umbral spikes may be the unresolved active component of the sunspots umbra discussed above. They represent finely structured shocked plasma showing up- and down-flows of order of 5-7 km s$^{-1}$. The data seem to suggest that the spikes are associated with interiors of strong flux tubes, thus conforming the idea about the existence of narrow channels conduct photospheric oscillations into the chromosphere \citep{2000Sci...288.1396S, 2005ApJ...635..670C}. 

Finally, \cite{2013ApJ...776...56R} suggested that the fine structure of UFs is related to the sunspot dynamic fibrils. The new data presented here shows that the dark filaments reported inside the UF area are the vertical umbral spikes projected on the otherwise uniform and unstructured UFs.

Authors thank anonymous referee for valuable suggestions and criticism. This work was conducted as part of the effort of NASA's Living with a Star Focused Science Team ``Jets''. We thank BBSO observing and engineering staff for support and observations. This research was supported by NASA LWS NNX11AO73G and NSF AGS-1146896 grants. VYu acknowledges support from Korea Astronomy and Space Science Institute during his stay there, where a part of the work was performed.
 

\end{document}